\documentstyle[11pt,newpasp,twoside,epsf]{article}
\def\edcomment#1{\iffalse\marginpar{\raggedright\sl#1\/}\else\relax\fi}
\reversemarginpar

\newcommand{\kmsmpc}{\kms\;{\rm Mpc}^{-1}}

\newcommand{\hmpc}{h^{-1}{\rm Mpc}}

\newcommand{\sgal}{\sigma_{\rm gal}}
\newcommand{\sdm}{\sigma_{\rm DM}}
\newcommand{\kms}{\;{\rm km}\,{\rm s}^{-1}}
\newcommand{\PSbox}[3]{\mbox{\rule{0in}{#3}\includegraphics{#1}\hspace{#2}}}

\begin{document}
\title{Group Scaling Relations From a Cosmological Hydrodynamic Simulation: No Pre-heating Required?}
\author{Romeel Dav\'e$^1$, Neal Katz$^2$, Lars Hernquist$^3$, David Weinberg$^4$}
\affil{$^1$ Steward Observatory, 933 N. Cherry Ave., Tucson, AZ 85721\\
$^2$ Astronomy Dept., Univ. of Massachussetts, Amherst, MA 01003\\
$^3$ Harvard/Smithsonian Center for Astrophyiscs, Cambridge, MA 02138\\
$^4$ Astronomy Dept., Ohio State Univ., Columbus, OH 43210}

\begin{abstract}

We investigate the X-ray vs. optical scaling relations of poor groups to
small clusters ($\sigma\approx 100-700$~km/s) identified in a cosmological
hydrodynamic simulation of a $\Lambda$CDM universe, with cooling and
star formation but no pre-heating.  We find that the scaling relations
between X-ray luminosity, X-ray temperature, and velocity dispersion
show significant departures from the relations predicted by simple
hydrostatic equilibrium models or simulations without cooling, having
steeper $L_X-\sigma$ and $L_X-T_X$ slopes and a ``break" at $\approx
200$~km/s ($\approx 0.3$~keV).  These departures arise because the
hot (X-ray emitting) gas fraction varies substantially with halo mass
in this regime.  Our predictions roughly agree with observations.
Thus radiative cooling is a critical physical process in modeling galaxy
groups, and may present an alternative to {\it ad hoc} models such as
pre-heating or entropy floors for explaining X-ray group scaling relations.
\end{abstract}

\section{Introduction}

The simplest view of a cluster is as a sphere of Virial-temperature
gas punctuated by old galaxies, with perhaps a cooling flow onto the cD
galaxy.  Within this model, the gas cooling time is longer than a Hubble
time everywhere except near the center, and thus it is predicted that
clusters should follow simple ``self-similar" scaling relations derived
from the Virial theorem combined with an assumption of hydrostatic
equilibrium, namely: $L_X\propto T_X^2$, $L_X\propto \sgal^4$, and
$T_X\propto \sgal^2$, where $L_X$ is the total X-ray luminosity, $T_X$
is the gas temperature (presumed to be the halo Virial temperature),
and $\sgal$ is the velocity dispersion of cluster galaxies (presumed to
trace the system's total mass).

While observations of the most massive clusters follow these relations,
as one progresses to smaller systems, the self-similar model fails.
The failure is most striking when one extends observations to poor groups,
where Helsdon \& Ponman (2000; HP) found that the luminosity of their poorest
groups declined much faster than expected ($L_X\propto T_X^{4.9}$), a
fact also interpreted as an ``entropy floor" of $\sim 100$~keV~cm$^{-2}$
(Ponman, Cannon \& Navarro 1999).  One model that successfully accounts
for these observations postulates that the gas is ``pre-heated" by some
unknown physical process (possibly supernovae or winds) with an energy
of $\sim 1$~keV/baryon.  However, the success of this model relies on
the assumption that groups are self-similarly scaled-down versions of
large clusters.

In these proceedings we present a preliminary test of whether self-similar
scaling relations arise naturally in a cosmological hydrodynamic
simulation of galaxy formation.  We find that on group scales, the
self-similar scaling relations are not followed, and that features can
arise that quantitatively mimic a pre-heating model.  The breaking of
self-similarity from clusters to groups is due to an increased cooling
efficiency in smaller systems, and does not imply additional heat or
entropy input at early times (see also Bryan 2000).

\section{Simulation and Group Identification}

Our simulation is of a $50\hmpc$ random volume in a $\Lambda$CDM universe
with $\Omega_m=0.4$, $H_0=65\kmsmpc$, and $\sigma_8=0.8$.  There are
$144^3$ dark matter and $144^3$ SPH particles.  It includes cooling,
star formation, and thermal feedback.  We identify galaxies using SKID,
and halos using a spherical-overdensity criterion on friends-of-friends
halos, as described in Murali et al. (2001).  Our 64-particle galaxy
mass resolution limit ($5\times 10^{10} M_\odot$) corresponds to $\approx
L_\star/4$.  Halos that contain three or more galaxies are identified as
``groups"; we find 128 at $z=0$.

\begin{figure}
\PSbox{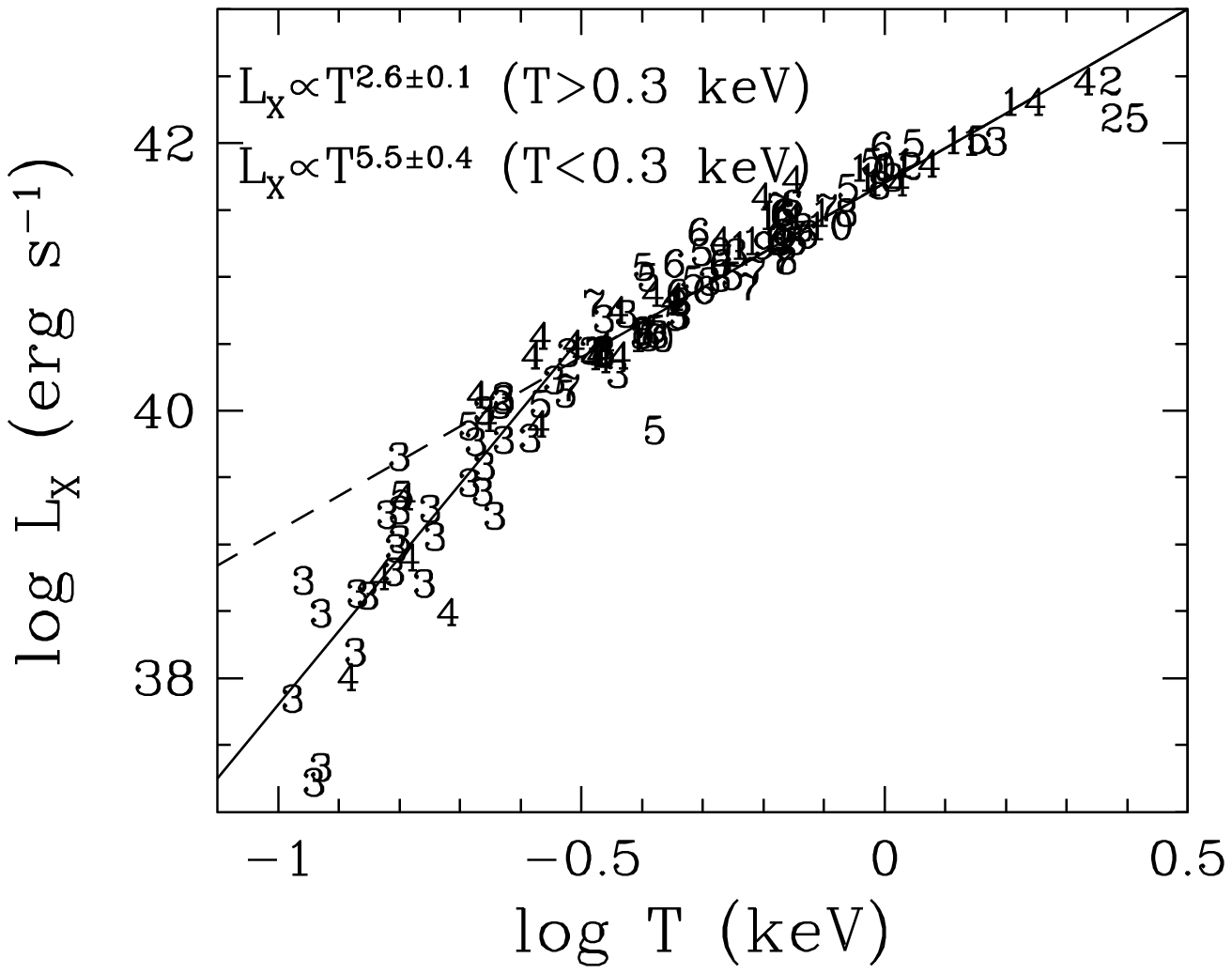 angle=0 voffset=-200 hoffset=50 vscale=48 hscale=48}{3.0in}{2.0in}
{\\\small Figure 1: $L_X-T_X$ relation, showing a break at $T\approx 0.3$~keV.}
\end{figure}

\begin{figure}
\PSbox{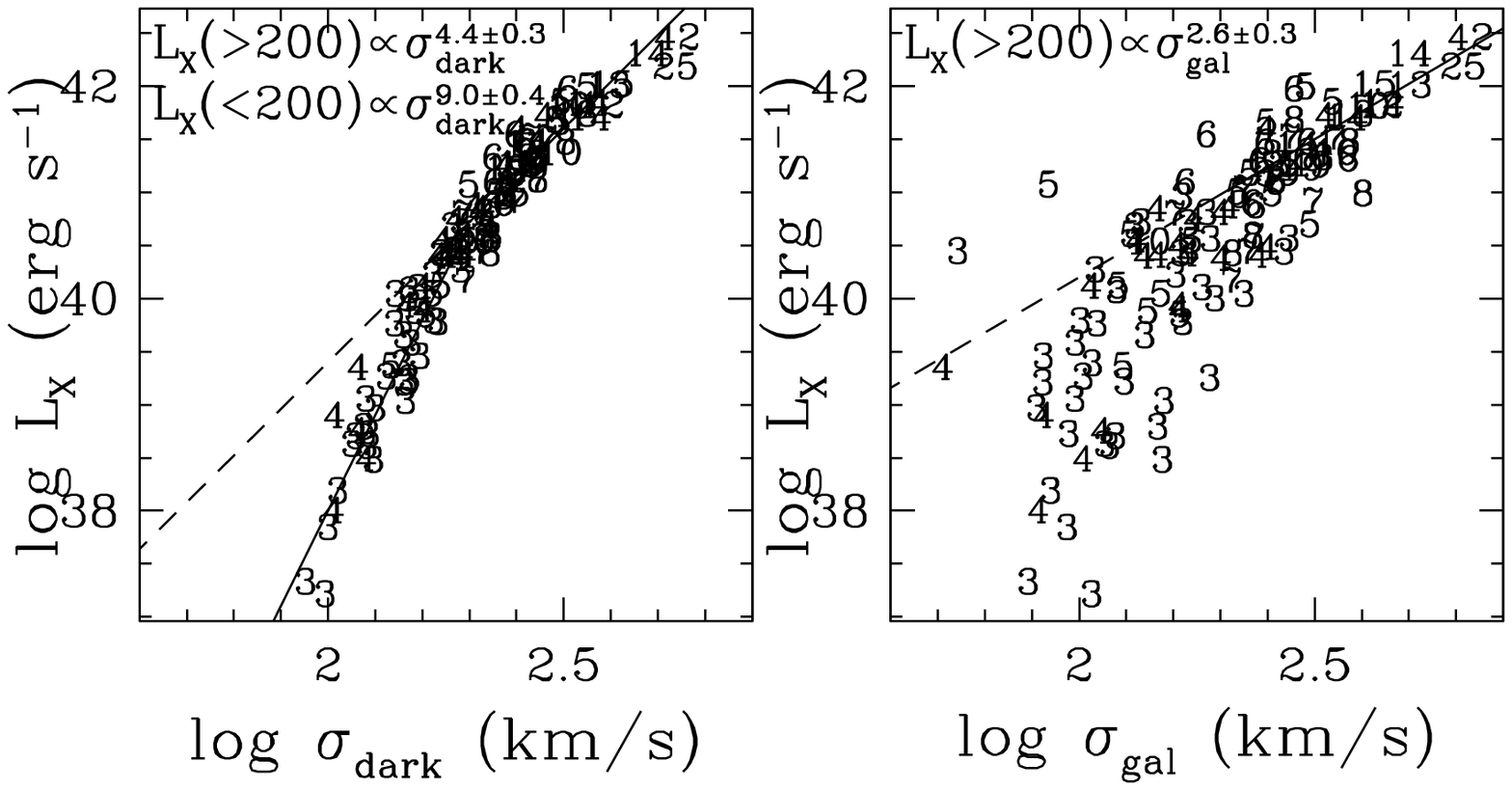 angle=0 voffset=-220 hoffset=50 vscale=48 hscale=48}{3.0in}{1.7in}
{\\\small Figure 2: $L_X-\sdm$ (left) and $L_X-\sgal$ (right) relations.}
\end{figure}

The X-ray luminosity of a group is computed from all particles with
$T>10^{4.5}$K that are members of that group.  Only Bremmstrahlung emission
is used; metal emission is not included in this preliminary analysis
to facilitate a straightforward comparison with the self-similar model,
and also because intragroup gas metallicities are uncertain.  In order to
avoid the well-known problem of overestimation of X-ray luminosity due to
SPH oversmoothing, we recalculate gas densities using only hot particles,
ignoring all particles with $T<10^{4.5}$K.  This explicit decoupling of the
hot and cold phases is shown to reproduce the correct X-ray luminosity
in analytic cases (Croft et al. 2001).  The X-ray temperature is given
by the average luminosity-weighted temperature of group particles.

\section{Results}

Figure~1 shows the $L_X-T_X$ relation for our simulated groups.
The plot symbols indicate the number of group members.  Best fit power
law relations above and below $T=0.3$~keV are shown in the upper left.
The slope, for groups with $T_X>0.3$~keV, is steeper than predicted by
the self-similar model, and steepens even more below $\approx 0.3$~keV
(the dashed line shows a continuation of the high-temperature slope).
The latter slope is consistent with the HP data, and the former with
results from clusters (White, Jones \& Forman 1997).  The break appears
at a somewhat lower temperature than is suggested by HP, but more
detailed modeling (i.e. including metal emission and aperture effects)
may improve agreement.  The crucial point is that a break occurs at all,
when no physics has been input into the simulation that specifically
picks out this temperature scale.  This suggests that a straightforward
extrapolation of cluster scaling relations (i.e. self-similarity) is an
inappropriate model for poor groups.

\begin{figure}
\PSbox{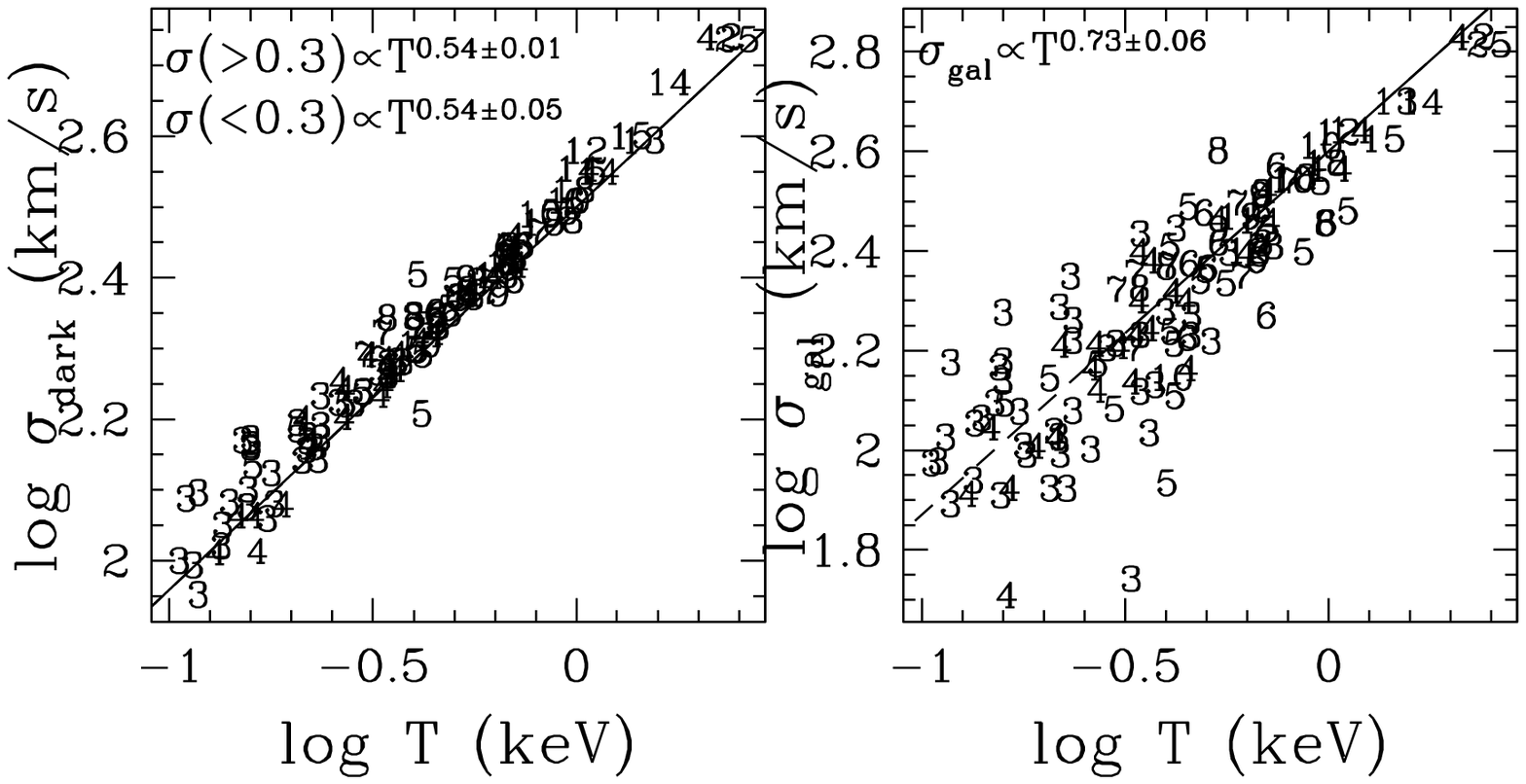 angle=0 voffset=-220 hoffset=50 vscale=48 hscale=48}{3.0in}{1.7in}
{\\\small Figure 3: $T_X-\sdm$ (left) and $T_X-\sgal$ (right) relations.}
\end{figure}

Figure~2 shows the $L_X-\sdm$ (left panel) and $\L_X-\sgal$ relations
(right panel).  The dark matter velocity dispersion faithfully traces the
group mass.  A break is seen in $L_X$ at $\sdm\approx 200$~km/s.
The slopes above and below the break are consistent with a compilation
of $L_X-\sigma$ for clusters down to galaxies by Mahdavi \& Geller (2001).
Their break occurs at $\sigma\approx 350$~km/s, but given observational
uncertainties it is roughly consistent with our prediction.

The galaxy velocity dispersion does {\it not} trace the group mass
when small numbers of galaxies are used.  The right panel of Figure~2
shows smaller groups shifted towards lower $\sgal$ compared to the left
panel, and the best-fit slope is altered significantly.  Thus we confirm
Zabludoff \& Mulchaey's (1998) result that a large number of galaxies
($\ga 10$) must be used to accurately estimate group velocity dispersions.
HP find a slope consistent with our $L_X-\sdm$, but claim that their
$\sigma$ is lowered due to misestimation from using small numbers of
galaxies.  However, Zimer, Zabludoff \& Mulchaey (these proceedings)
find a similar slope using more than 20 galaxies per group.

Figure~3 shows the $T_X-\sigma$ relation for $\sdm$ (left panel) and
$\sgal$ (right panel).  Interestingly, the $T_X-\sdm$ relation shows no
break at 200~km/s, indicating that the drop in $L_X$ is not due to an
additional heat source such as shock heating on filaments.  The slope
is in reasonable agreement with the self-similar model, indicating that
these systems are Virialized.  However, the misestimation of $\sigma$
from using too few galaxies results in a steeper slope (right panel),
bringing it more into agreement with HP; this can be misinterpreted as
excess heat injection.

Since there is no excess heat, we conclude that the break in $L_X$
arises from a drop in the hot fraction due to an increase in efficiency
of galaxy formation in smaller systems.  The fraction of hot ($T>10^5$~K)
gas for our simulated groups drops linearly from 50\% at $\sigma>500$~km/s
to 20\% at $\sigma\approx 100$~km/s.  When placed on a logarithmic scale
(cf. the $L_X$ plots), the decline appears to steepen at lower $\sigma$;
hence the appearance of a ``break".  While the large cold gas + stellar
fraction may appear to contradict baryon estimates in clusters, there
are indications that the hot gas fraction decreases in smaller systems
(see Bryan 2000 for summary).  In any case, the trend for a rapid decline
of the hot fraction in this mass regime seems to be a generic feature
of CDM-based galaxy formation, arising in semi-analytic models as well
(Bower, these proceedings).

\section{Summary}

Using a cosmological hydrodynamic simulation, we have shown that
group X-ray scaling relations are not expected to follow simple
extrapolations from clusters.  For rich groups, the scaling relations
for the X-ray luminosity are somewhat different than predicted by
self-similarity arguments, while for poor groups the scaling relations
deviate dramatically.  The trends of our predictions are in agreement
with observations.  The location of the predicted break predicted is
slightly different than that observed, but more careful modeling and
better data may improve the agreement.  We conclude that cooling is
responsible for breaking self-similarity, and hence it is premature
to use observed poor group scaling relations as evidence for an excess
injection of heat or entropy into these systems.

\end{document}